
\font\twelvebf=cmbx10 scaled \magstep1
\hsize 15.2true cm
\vsize 22.0true cm
\voffset 1.5cm
\nopagenumbers
\headline={\ifnum \pageno=1 \hfil \else\hss\tenrm\folio\hss\fi}
\pageno=1
 
\hfill IUHET--345, IU/NTC--96--11, DFTT 65/96
 
\hfill Nov. 1996
\bigskip
\noindent{\twelvebf Diquark model of dibaryons$^1$}
 
\medskip
\bigskip
 
\noindent D. B. Lichtenberg$^a$ and Renato Roncaglia$^b$
 
\noindent Physics Department, Indiana University, 
Bloomington, IN 47405, USA

\noindent Enrico Predazzi$^c$ 

\noindent Dipartimento di Fisica Teorica, 
Universit\`a di Torino
 and INFN, Sezione di Torino, I-10125,
Torino, Italy
 
 \vskip 1 cm

\noindent {\bf Abstract.} A diquark model previously
formulated  to describe exotic mesons is extended
to dibaryons. In the model, dibaryons containing only
light quarks are unbound. The $H$ dibaryon,
consisting of $uuddss$ quarks, is unstable by about 
90 MeV or more, and should 
decay strongly into two $\Lambda$ baryons. 
A charmed dibaryon $H_c$, composed of $uuddsc$, is
unstable by about 60 MeV or more 
and should decay strongly into $\Lambda + \Lambda_c$. 
On the other hand, we find that a bottom dibaryon 
$H_b$, made of $uuddsb$, may be just bound by about 10 MeV
with respect to $\Lambda + \Lambda_b$. If so, it
should decay weakly into  various final states with
a charmed hadron. A possible two-body decay
would be into $\Lambda + \Lambda_c$. 

\bigskip
\medskip
\bigskip

\vskip 6cm
$^1$Invited talk given by the first
author at an international workshop, 
Diquarks 3, Torino, Italy, October 28--30, 1996. A slightly
different version will appear in the proceedings, to be
published by World Scientific, Singapore.
\medskip
$^a$lichten@indiana.edu
 
$^b$roncagli@indiana.edu
 
$^c$predazzi@to.infn.it

\vfill\eject
 
\bigskip 

The deuteron is the only known dibaryon. However, many
authors have proposed that there might exist others,
some of them not well-separated resonant states of two 
nucleons but
rather six-quark states in other configurations. 
A recent review [1]
contains some references to the subject.

  From time to time experiments have indicated 
that dibaryons other than the deuteron
might exist, but none of these experiments has been 
decisive. One  work reporting a possible dibaryon
signal in a system containing only light $u$ and $d$
quarks appeared very recently [2], Another paper
reporting evidence for a dibaryon resonance with 
strangeness $-1$ appeared several years earlier [3].
 
Here, we concentrate on the $H$ dibaryon, 
composed of $uuddss$ quarks, which Jaffe [4] proposed might
be stable against decay into two $\Lambda$ baryons. We
also briefly discuss the $H_c$, in which one of the
$s$ quarks is replaced by a $c$ quark, and the $H_b$,
in which the $c$ is replaced by a $b$. 
Thus far, the $H$ dibaryon has not been observed, and
theorists have disagreed about
whether it is bound or not. Oka [5] has given
a recent review of the theory of the $H$. 
The $H_c$ and $H_b$, which have
received much less attention, have also not been observed.

In general, we are interested in the lowest-mass dibaryon 
states with
a given quark content, as these are most likely to be
observed. We can envision two different types of 
stable or nearly stable
dibaryons. The first, like the deuteron, can be considered
as a kind of 
``molecule'' composed of two largely non-overlapping
colour-singlet
baryons, and the second is composed of six quarks, no
three of which form a colour singlet. Thus far, the
only known example of the first kind of dibaryon is
the deuteron, and there is no 
definitely known example of the second
kind. In this work, we consider only the second kind
of dibaryon. 

One reason to consider a model for dibaryons is that,
so far, lattice gauge calculations have not been
adequate to decide on the nature of the $H$, one
calculation finding that the $H$ is unbound [6], and
another finding the opposite [7].

In our model, a dibaryon is composed of three diquarks,
each of which is a colour antitriplet. In the $H$, 
$H_c$, and $H_b$  each diquark has spin zero. 
Two of the diquarks couple to form a colour-triplet
quadriquark, which, in turn couples to the third
diquark to form a colour singlet. We shall often refer
to a diquark as a colour triplet, although, of course,
like an antiquark,  it is a colour antitriplet.

There are two reasons why we confine ourselves to
colour-triplet diquarks. (1) At
small interquark separations, the force between 
two quarks arising from QCD
is attractive in a triplet state,
while it is repulsive in a sextet state. This fact
makes it plausible that the mass of a colour-sextet
diquark is considerably larger than that of a 
colour-triplet diquark, leading to a larger dibaryon
mass, and consequently to a more unstable dibaryon
that will be harder to observe. (2) With plausible 
assumptions motivated by
QCD, we are able to obtain the masses of  colour-triplet
diquarks and the masses of dibaryons just from the
properties of mesons and baryons,
without the need for an explicit Hamiltonian. We have
already applied these ideas to possible exotic mesons,
each of which is composed of a 
diquark and antidiquark [8].

Our diquarks are constructed so that the two quarks
in each diquark satisfy the Pauli principle. 
In the limit that diquarks are pointlike,
or at least much smaller than the typical separation
between different diquarks, we can safely ignore the Pauli 
principle for quarks in different diquarks. This is
analogous to the case in nuclear physics, in which the quark 
degrees of freedom in nucleons are generally ignored.

In nuclei, when two nucleons are close
together, an effective repulsive potential exists between 
them that arises in part because of the Pauli principle
for quarks. Similarly, when two diquarks in a dibaryon
are close together, the Pauli principle
between quarks of the same flavour in different
diquarks leads to a
modification of the effective potential between the
diquarks. This modification should cause a short-range
repulsion. We neglect this repulsion, and thereby 
underestimate the mass of a dibaryon containing 
three diquarks if  each of two different diquarks contains
a quark of the same flavour.  To lessen this effect,
we restrict ourselves to dibaryons containing three
different diquarks and at most two quarks of any
flavour. Furthermore, we restrict each diquark 
to have two different flavoured quarks. 

The restriction that each diquark has two different quarks
enables us to consider diquarks with spin zero. A spin-zero
diquark has a lower mass than a spin-one diquark with
the same quark content
because of the colour-magnetic interaction. 
Consequently, a dibaryon made of spin-zero diquarks
is likely to have a lower mass than a dibaryon
made of spin-one diquarks, making the former more easily 
observable.

The masses of the diquarks have already been calculated
(see Table 1 of Ref.\ [8])
in terms of properties of mesons and baryons, and
we simply take over the results. We then combine 
two diquarks into a colour-triplet quadriquark.
We can obtain the mass of the quadriquark by adding the
appropriate interaction energy $E^t_{12}$, 
where the superscript $t$ denotes the fact that
the quadriquark is a colour triplet. The interaction 
energy can be computed for any reduced mass by interpolation
or extrapolation by the methods of Ref.\ [8].
Once we have the
mass of the quadriquark, we can obtain the reduced
mass of it and the remaining diquark and so obtain
the appropriate interaction energy $E_{12}$. 
Of all possible ways to choose the quadriquark from
two out of three diquarks, we pick the way that yields
the lowest dibaryon mass, as we are looking for a
lower limit.
We now have all the ingredients to obtain 
lower limits on the masses
of dibaryons from the diquark masses of Ref.\ [8] and
the interaction energies $E^t_{12}$ and $E_{12}$.

We obtain that the mass of the $H$ dibaryon is
$$M(H) \geq  2320  \ {\rm MeV}, \eqno(1)$$
rounded to the nearest 10 MeV.
This mass is about 90 MeV above the mass of two
$\Lambda $ baryons, and so we obtain that the $H$
can decay strongly into two $\Lambda$'s. Because
in our model there is nothing to inhibit the decay,
the decay width may be too large to make the state
readily observable. 
Similarly, we obtain that the mass of the $H_c$ dibaryon
is
$$M(H_c) \geq 3460 \ {\rm MeV}, \eqno(2)$$
a value  60 MeV above threshold for decay into
$\Lambda + \Lambda_c$. Again, the decay width might
be too large to allow the state to be observable.
Within our model,
we estimate errors on the lower limits in (1) and
(2) to be about
30 MeV, owing to the methods used to calculate the
diquark masses. 
Lastly, we obtain that the mass of the $H_b$ is
$$M(H_b) \geq 6730 \ {\rm MeV}, \eqno(3)$$
It is interesting that this last value is $10\pm 20$ MeV
{\it below} the threshold for decay into
$\Lambda + \Lambda_b$. Because our method allows
us only to calculate a lower limit on the mass of
the $H_b$, we cannot say definitely whether it
is bound. However, this dibaryon is well worth
searching for. If it is bound, it should decay weakly
into the two-body final state $\Lambda + \Lambda_c$ as
well as into other hadrons including a charmed hadron. 

We estimate that dibaryons containing 
diquarks with only $u$ and $d$
quarks  will be even more unstable than the $H$. 
We conclude that, according to our model 
in which dibaryons
are composed of colour-triplet diquarks, dibaryons
containing at most one $c$ quark and lighter quarks
will be unstable against strong decay and hard to observe,
but a dibaryon containing one $b$ quark might 
be stable against strong decay. 

Our model of a dibaryon is that it is composed of three
diquarks, each of which is a colour triplet. 
Within our model, we have made approximations
that lower the calculated dibaryon mass compared to 
its mass in an exact model calculation. This means
that our calculated masses should be regarded as 
lower limits. Consequently,
if a weakly decaying
$H$ or $H_c$ dibaryon is observed, it must have a structure
quite different from the one we have assumed.

\bigskip
Part of this work was done while one of us (E.P.) visited
Indiana University.  This work was supported in
part by the U.S. Department of Energy, by the Italian
Institute for Nuclear Physics (INFN) and
by the Ministry of Universities, Research,
Science and Technology (MURST) of Italy.

\vfill\eject

\bigskip
\noindent References
\medskip
[1] M. Anselmino, E. Predazzi, S. Ekelin,
S. Fredriksson, and D.B. Lichtenberg, Rev. Mod. Phys.
{\bf 65}, 1199 (1993).

[2] W. Brodowski et al., Z. Phys.\ A {\bf 355} (1996) 5.

[3] H. Pickarz, in {\it Intersections Between Particle
and Nuclear Physics}, AIP Conf. Proc. 243, editor, 
W.T.H. Van Oers, AIP New York (1992). 

[4] R.L. Jaffe, Phys. Rev. Lett. 38 (1977) 195.

[5] M. Oka, in Intern. Symposium on Exotic Atoms and
Nuclei, June 7--10, 1995, Hakono, Japan. Proceedings
to be published in Hyperfine Interactions, R. Hayano, ed.

[6] P.B. Mackenzie and H.B. Thacker, Phys.\ Rev.\
Lett. {\bf 55} (1985) 2539.

[7] Y.T. Iwasaki, T. Yoshie, and Y. Tsuboi, Phys.\ Rev.\
Lett. {\bf 60} (1988) 1371.

[8] D.B. Lichtenberg, R. Roncaglia, and E. Predazzi,
Diquark model of exotic mesons, previous talk given at the 
Diquark III workshop (Torino, Oct. 28--30, 1996)

\bye